\documentclass[conference]{IEEEtran}
\IEEEoverridecommandlockouts
\usepackage{cite}
\usepackage{amsmath,amssymb,amsfonts}
\usepackage{algorithm}
\usepackage{subcaption}
\usepackage[noend]{algpseudocode}
\usepackage{graphicx}
\usepackage{textcomp}
\usepackage{xcolor}
\usepackage{pgf, tikz}
\usepackage[frozencache,cachedir=.]{minted}
\usepackage{booktabs}
\usepackage{multirow}
\usepackage{circuitikz}
\usepackage{dblfloatfix}    

\usepackage{pgfplots}

\usetikzlibrary{automata,patterns,calc,shapes.multipart,shapes.symbols,arrows,matrix,arrows.meta,chains,decorations.pathreplacing,fit,shadows}
\makeatletter
\def\BState{\State\hskip-\ALG@thistlm}
\makeatother

\makeatletter

\pgfdeclareshape{dff}{
  \savedanchor\northeast{%
    \pgfmathsetlength\pgf@x{\pgfshapeminwidth}%
    \pgfmathsetlength\pgf@y{\pgfshapeminheight}%
    \pgf@x=0.4\pgf@x
    \pgf@y=0.4\pgf@y
  }
  \savedanchor\southwest{%
    \pgfmathsetlength\pgf@x{\pgfshapeminwidth}%
    \pgfmathsetlength\pgf@y{\pgfshapeminheight}%
    \pgf@x=-0.4\pgf@x
    \pgf@y=-0.4\pgf@y
  }
  \inheritanchorborder[from=rectangle]

  \anchor{center}{\pgfpointorigin}
  \anchor{north}{\northeast \pgf@x=0pt}
  \anchor{east}{\northeast \pgf@y=0pt}
  \anchor{south}{\southwest \pgf@x=0pt}
  \anchor{west}{\southwest \pgf@y=0pt}
  \anchor{north east}{\northeast}
  \anchor{north west}{\northeast \pgf@x=-\pgf@x}
  \anchor{south west}{\southwest}
  \anchor{south east}{\southwest \pgf@x=-\pgf@x}
  \anchor{text}{
    \pgfpointorigin
    \advance\pgf@x by -.5\wd\pgfnodeparttextbox%
    \advance\pgf@y by -.5\ht\pgfnodeparttextbox%
    \advance\pgf@y by +.5\dp\pgfnodeparttextbox%
  }

  \anchor{D}{
    \pgf@process{\northeast}%
    \pgf@x=-1\pgf@x%
    \pgf@y=.5\pgf@y%
  }
  \anchor{CLK}{
    \pgf@process{\northeast}%
    \pgf@x=-1\pgf@x%
    \pgf@y=-.66666\pgf@y%
  }
  \anchor{Q}{
    \pgf@process{\northeast}%
    \pgf@y=.5\pgf@y%
  }
  \anchor{Qn}{
    \pgf@process{\northeast}%
    \pgf@y=-.5\pgf@y%
  }
  \backgroundpath{
    \pgfpathrectanglecorners{\southwest}{\northeast}
    \pgf@anchor@dff@CLK
    \pgf@xa=\pgf@x \pgf@ya=\pgf@y
    \pgf@xb=\pgf@x \pgf@yb=\pgf@y
    \pgf@xc=\pgf@x \pgf@yc=\pgf@y
    \pgfmathsetlength\pgf@x{1.6ex} 
    \advance\pgf@ya by \pgf@x
    \advance\pgf@xb by \pgf@x
    \advance\pgf@yc by -\pgf@x
    \pgfpathmoveto{\pgfpoint{\pgf@xa}{\pgf@ya}}
    \pgfpathlineto{\pgfpoint{\pgf@xb}{\pgf@yb}}
    \pgfpathlineto{\pgfpoint{\pgf@xc}{\pgf@yc}}
    \pgfclosepath

    \begingroup
    \tikzset{flip flop/port labels} 
    \tikz@textfont

    \pgf@anchor@dff@D
    \pgftext[left,base,at={\pgfpoint{\pgf@x}{\pgf@y}},x=\pgfshapeinnerxsep]{\raisebox{-0.75ex}{D}}

    \pgf@anchor@dff@Q
    \pgftext[right,base,at={\pgfpoint{\pgf@x}{\pgf@y}},x=-\pgfshapeinnerxsep]{\raisebox{-.75ex}{Q}}

    \pgf@anchor@dff@Qn
    \pgftext[right,base,at={\pgfpoint{\pgf@x}{\pgf@y}},x=-\pgfshapeinnerxsep]{\raisebox{-.75ex}{$\overline{\mbox{Q}}$}}
    
    \endgroup
  }
}

\tikzset{add font/.code={\expandafter\def\expandafter\tikz@textfont\expandafter{\tikz@textfont#1}}} 

\tikzset{flip flop/port labels/.style={font=\scriptsize}}
\tikzset{every dff node/.style={draw,minimum width=2cm,minimum 
height=2.828427125cm,very thick,inner sep=1mm,outer sep=0pt,cap=round}}

\makeatother

\begin{document}
\title{\texttt{vlang}: Mapping Verilog Netlists to Modern Technologies}
\author{\IEEEauthorblockN{Nicholas V. Giamblanco}
\IEEEauthorblockA{\textit{Computer Engineering Department} \\
\textit{University of Toronto}\\
Toronto, Canada \\
\texttt{giambla2@ece.utoronto.ca}}
\and
\IEEEauthorblockN{Andrew Schmidt}
\IEEEauthorblockA{\textit{Information Sciences Institute} \\
\textit{University of Southern California}\\
Arlington, USA \\
\texttt{aschmidt@isi.edu}}
\thanks{This work is funded by the Department of Energy's Kansas City National Security Campus, operated by Honeywell Federal Manufacturing \& Technologies, LLC, under contract number DE-NA0002839.}
}

\maketitle

\begin{abstract}
 Portability of hardware designs between Programmable Logic Devices (PLD) can be accomplished through the use of device-agnostic hardware description languages (HDL) such as Verilog or VHDL. Hardware designers can use HDLs to migrate hardware designs between devices and explore performance, area and power tradeoffs, as well as, port designs to an alternative device. However, if design files are corrupt or missing, the portability of the design is lost. While reverse engineering efforts may be able to recover an HDL-netlist of the original design, HDL-netlists use device-specific primitives, restricting portability. Additionally, the recovered design may benefit from other computational technologies (e.g., $\mu$P, GPGPUs), but is restricted to the domain of PLDs. In this work, we provide a new framework, \texttt{vlang}, which automatically maps Verilog-netlists into LLVM's \textit{intermediate representation} (IR). The remapped design can use the LLVM-framework to target many device technologies such as: \texttt{x86-64} assembly, RISC-V, ARM or to other PLDs with a modern high-level synthesis tool. Our framework is able to preserve the exact functionality of the original design within the software executable. The \texttt{vlang}-produced software executable can be used with other software programs, or to verify the functionality and correctness of the remapped design. We evaluate our work with a suite of hardware designs from OpenCores. We compare our framework against state-of-the-art simulators, thereby outlining our framework's ability to produce a fully-functional, cycle accurate software-executable. We also explore the usage of \texttt{vlang} as a front-end for high-level synthesis tools.

\end{abstract}

\begin{IEEEkeywords}
Verilog-to-C Compiler, High-Level Synthesis, Design Portability, Graph Algorithms
\end{IEEEkeywords}

\section{Introduction} \label{sec:intro}

Programmable logic devices (PLDs) are rapidly evolving: many new devices technologies and programming models are being delivered to the market to meet the computing needs of today~\cite{eeckhout2017moore,ahmad2019xilinx,koeplinger2018spatial,chin2017cgra,li2019survey}. However, maintaining portability of designs between past, current or future PLD-technologies is crucial ~\cite{giamblanco2019asap, giamblanco2020dynamic}: either for continued product use, or for broad adoption of upcoming technologies. 

Traditionally, the portability of hardware designs is generally handled with a portable design medium such as a hardware-description language (HDL) or with specialized design files (i.e., bitstream files). However, if these design mediums are lost or damaged, preserving the design on a programmable device is difficult. If a PLD is still configured with the design's bitstream, it may be possible to recover the design through \textit{bitstream} readback \cite{benz2012bil,quadir2016survey,note2008bitstream,yu2018recent}. Additionally, the original PLD model may be deprecated (i.e., the PLD model is no longer manufactured), rendering the recovered bitstream as useless since it cannot be applied to any other PLD. Therefore, the bitstream will need to undergo reverse engineering efforts~\cite{note2008bitstream}.

Typically, the recovered design will be in the form of primitives native to the device (e.g., device-specific LUTs, DSPs, BRAMs, etc.), and represented with a HDL, which is referred to as a HDL netlist \cite{benz2012bil,note2008bitstream}. The primitives used in the HDL netlist may not exist on the desired replacement device. Therefore, additional methods must be applied to remap deprecated-device primitives in the HDL netlist to the fabric of a user-selected modern device. Once this design has been remapped to the new-device's primitives, it will undergo the corresponding vendor's compilation flow. 

It is not clear if the recovered-design will retain similar performance, power and area characteristics on alternative PLD technologies. Additionally, the recovered hardware design may benefit from other computational technologies (e.g. a $\mu$Processor, GPGPU, CGRA). However, the reverse engineering process restricts portability of the recovered design to only programmable-hardware platforms. This is due the structural encoding of the recovered design: the low-level representation of an algorithm or hardware design limits the possibility to explore design optimizations or modifications during the recovery process (i.e. security patches, performance improvements, power reduction techniques) or to gain an understanding of an application's intentions.  

To address the issue of hardware-design portability with HDL-netlists, we present a framework, \texttt{vlang}. Our framework is a new LLVM \cite{lattner2004llvm} front-end which compiles Verilog-HDL netlists into LLVM's intermediate representation (IR), permitting portability of designs across a variety of computational devices (e.g., \texttt{x86}, RISC-V, ARM, etc.). \texttt{vlang}'s LLVM-IR is high-level synthesis (HLS) compatible, allowing designs to be remapped to alternative PLD technologies supported by modern HLS tools. Additionally, \texttt{vlang} produces LLVM-IR code which is cycle-accurate and functionally equivalent and to a hardware circuit, and can be used for simulation. We demonstrate that we can compile \texttt{vlang}'s LLVM-IR into a software executable which can be linked into C/C++ programs.

\tikzset{
pattern size/.store in=\mcSize, 
pattern size = 5pt,
pattern thickness/.store in=\mcThickness, 
pattern thickness = 0.3pt,
pattern radius/.store in=\mcRadius, 
pattern radius = 1pt}
\makeatletter
\pgfutil@ifundefined{pgf@pattern@name@_acwuvktkv lines}{
\pgfdeclarepatternformonly[\mcThickness,\mcSize]{_acwuvktkv}
{\pgfqpoint{0pt}{0pt}}
{\pgfpoint{\mcSize+\mcThickness}{\mcSize+\mcThickness}}
{\pgfpoint{\mcSize}{\mcSize}}
{\pgfsetcolor{\tikz@pattern@color}
\pgfsetlinewidth{\mcThickness}
\pgfpathmoveto{\pgfpointorigin}
\pgfpathlineto{\pgfpoint{\mcSize}{0}}
\pgfusepath{stroke}}}
\makeatother
\tikzset{every picture/.style={line width=0.75pt}} 

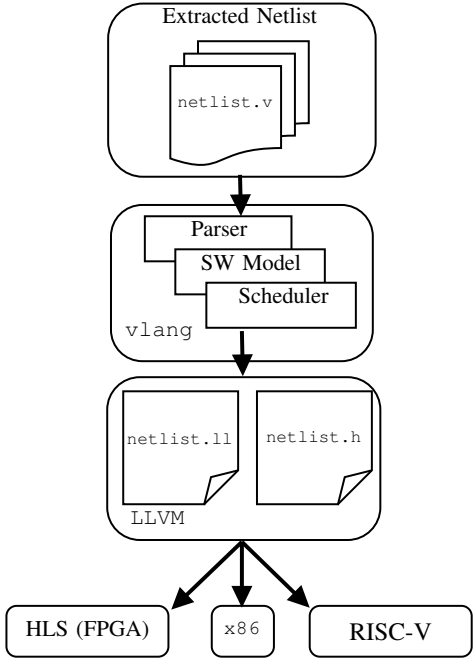
\begin{figure}[tbh!]
    \centering
\begin{tikzpicture}[x=0.75pt,y=0.75pt,yscale=-1,xscale=1,scale=0.95, every node/.style={scale=0.95}]

\draw  [fill={rgb, 255:red, 255; green, 255; blue, 255 }  ,fill opacity=1 ] (106.5,191.6) .. controls (106.5,182.43) and (113.93,175) .. (123.1,175) -- (228.9,175) .. controls (238.07,175) and (245.5,182.43) .. (245.5,191.6) -- (245.5,241.4) .. controls (245.5,250.57) and (238.07,258) .. (228.9,258) -- (123.1,258) .. controls (113.93,258) and (106.5,250.57) .. (106.5,241.4) -- cycle ;
\draw  [fill={rgb, 255:red, 255; green, 255; blue, 255 }  ,fill opacity=1 ] (152.7,88) -- (211.5,88) -- (211.5,131.56) .. controls (174.75,131.56) and (182.1,147.27) .. (152.7,137.1) -- cycle ; \draw  [fill={rgb, 255:red, 255; green, 255; blue, 255 }  ,fill opacity=1 ] (145.35,94.6) -- (204.15,94.6) -- (204.15,138.16) .. controls (167.4,138.16) and (174.75,153.87) .. (145.35,143.7) -- cycle ; \draw  [fill={rgb, 255:red, 255; green, 255; blue, 255 }  ,fill opacity=1 ] (138,101.2) -- (196.8,101.2) -- (196.8,144.76) .. controls (160.05,144.76) and (167.4,160.47) .. (138,150.3) -- cycle ;
\draw   (105,86.4) .. controls (105,76.24) and (113.24,68) .. (123.4,68) -- (228.6,68) .. controls (238.76,68) and (247,76.24) .. (247,86.4) -- (247,141.6) .. controls (247,151.76) and (238.76,160) .. (228.6,160) -- (123.4,160) .. controls (113.24,160) and (105,151.76) .. (105,141.6) -- cycle ;
\draw   (156,334) -- (113,334) -- (113,274) -- (174,274) -- (174,316) -- cycle -- (156,334) ; \draw   (174,316) -- (159.6,319.6) -- (156,334) ;
\draw  [fill={rgb, 255:red, 255; green, 255; blue, 255 }  ,fill opacity=1 ] (124.47,181) -- (202.32,181) -- (202.32,204.94) -- (124.47,204.94) -- cycle ;
\draw  [fill={rgb, 255:red, 255; green, 255; blue, 255 }  ,fill opacity=1 ] (140.66,198.56) -- (220.03,198.56) -- (220.03,222.5) -- (140.66,222.5) -- cycle ;
\draw  [fill={rgb, 255:red, 255; green, 255; blue, 255 }  ,fill opacity=1 ] (157.13,216.31) -- (236.5,216.31) -- (236.5,240.25) -- (157.13,240.25) -- cycle ;
\draw   (104.5,283.8) .. controls (104.5,274.25) and (112.25,266.5) .. (121.8,266.5) -- (232.2,266.5) .. controls (241.75,266.5) and (249.5,274.25) .. (249.5,283.8) -- (249.5,335.7) .. controls (249.5,345.25) and (241.75,353) .. (232.2,353) -- (121.8,353) .. controls (112.25,353) and (104.5,345.25) .. (104.5,335.7) -- cycle ;
\draw [line width=1.5]    (175,159) -- (175.41,177) ;
\draw [shift={(175.5,181)}, rotate = 268.7] [fill={rgb, 255:red, 0; green, 0; blue, 0 }  ][line width=0.08]  [draw opacity=0] (11.61,-5.58) -- (0,0) -- (11.61,5.58) -- cycle    ;

\draw [line width=1.5]    (176,242) -- (176.42,262) ;
\draw [shift={(176.5,266)}, rotate = 268.81] [fill={rgb, 255:red, 0; green, 0; blue, 0 }  ][line width=0.08]  [draw opacity=0] (11.61,-5.58) -- (0,0) -- (11.61,5.58) -- cycle    ;

\draw   (227,334) -- (184,334) -- (184,274) -- (245,274) -- (245,316) -- cycle -- (227,334) ; \draw   (245,316) -- (230.6,319.6) -- (227,334) ;
\draw [line width=1.5]    (176,354) -- (141.39,387.23) ;
\draw [shift={(138.5,390)}, rotate = 316.16999999999996] [fill={rgb, 255:red, 0; green, 0; blue, 0 }  ][line width=0.08]  [draw opacity=0] (11.61,-5.58) -- (0,0) -- (11.61,5.58) -- cycle    ;

\draw   (51,393.4) .. controls (51,390.42) and (53.42,388) .. (56.4,388) -- (131.6,388) .. controls (134.58,388) and (137,390.42) .. (137,393.4) -- (137,409.6) .. controls (137,412.58) and (134.58,415) .. (131.6,415) -- (56.4,415) .. controls (53.42,415) and (51,412.58) .. (51,409.6) -- cycle ;
\draw [line width=1.5]    (176,354) -- (176,385) ;
\draw [shift={(176,389)}, rotate = 270] [fill={rgb, 255:red, 0; green, 0; blue, 0 }  ][line width=0.08]  [draw opacity=0] (11.61,-5.58) -- (0,0) -- (11.61,5.58) -- cycle    ;

\draw   (160,393.4) .. controls (160,390.42) and (162.42,388) .. (165.4,388) -- (187.6,388) .. controls (190.58,388) and (193,390.42) .. (193,393.4) -- (193,409.6) .. controls (193,412.58) and (190.58,415) .. (187.6,415) -- (165.4,415) .. controls (162.42,415) and (160,412.58) .. (160,409.6) -- cycle ;
\draw [line width=1.5]    (176,354) -- (209.13,386.21) ;
\draw [shift={(212,389)}, rotate = 224.19] [fill={rgb, 255:red, 0; green, 0; blue, 0 }  ][line width=0.08]  [draw opacity=0] (11.61,-5.58) -- (0,0) -- (11.61,5.58) -- cycle    ;

\draw   (212,393.4) .. controls (212,390.42) and (214.42,388) .. (217.4,388) -- (292.6,388) .. controls (295.58,388) and (298,390.42) .. (298,393.4) -- (298,409.6) .. controls (298,412.58) and (295.58,415) .. (292.6,415) -- (217.4,415) .. controls (214.42,415) and (212,412.58) .. (212,409.6) -- cycle ;

\draw (167,120) node   [align=left] {{\fontfamily{pcr}\selectfont {\scriptsize netlist.v}}};
\draw (143,299) node   [align=left] {{\fontfamily{pcr}\selectfont {\scriptsize netlist.ll}}};
\draw (214,298) node   [align=left] {{\fontfamily{pcr}\selectfont {\scriptsize netlist.h}}};
\draw (161.69,187.58) node   [align=left] {{ {\small Parser}}};
\draw (178.64,205.35) node   [align=left] {{ {\small SW Model}}};
\draw (195.87,223.1) node   [align=left] {{ {\small Scheduler}}};
\draw (174.87,74.1) node   [align=left] {{\small Extracted Netlist}};
\draw (132,242) node   [align=left] {{\small {\fontfamily{pcr}\selectfont vlang}}};
\draw (131,341) node   [align=left] {{\fontfamily{pcr}\selectfont {\small LLVM}}};
\draw (94,401) node   [align=left] {{\small HLS (FPGA)}};
\draw (176,400) node   [align=left] {{\fontfamily{pcr}\selectfont {\footnotesize x86}}};
\draw (255,401.5) node   [align=left] {RISC-V};

\end{tikzpicture} 
    \caption{An Overview of \texttt{vlang}'s compilation process. }
    \label{fig:vlang}
\end{figure}

The main contributions of this paper are:
\begin{itemize}
    \item A LLVM-frontend, \texttt{vlang}, which compiles a Verilog-netlist into LLVM-IR program which is functionally equivalent and cycle accurate to the Verilog-netlist.
    \item A performance comparison of \texttt{vlang}'s simulation capabilities against state-of-the-art tools.
    \item A study on \texttt{vlang}'s ability to preserve the functionality of designs through the use of modern HLS tools.
\end{itemize}

\section{Background}

We provide: (1) a brief review of the LLVM compilation framework and (2) device-specific netlist example to assist the reader in understanding our work. 

\subsection{LLVM}
The LLVM compilation framework \cite{lattner2004llvm} provides a set of toolchains to enable compilation between any supported source-language to any supported target language. Source-languages are compiled into LLVM's-IR, a strongly-typed RISC instruction-set. The IR is agnostic to the source and target languages. RISC instructions are simple operations such as \texttt{add}, \texttt{or}, \texttt{xor}, \texttt{store}, \texttt{load}, etc. Instructions may have: (a) 0 or more inputs and (b) 0 or 1 outputs. LLVM encodes the IR using single static assignment \cite{leung1999static}. Control flow in the LLVM-IR is handled with the use of basic blocks and control flow graphs. A basic-block requires an entry to the block, a set of control-flow-independent instructions, and an termination point. Termination points can either be branch (\texttt{br}) or return (\texttt{ret}) instructions. Connections between basic-blocks are made if a branch exists to some other basic-block or itself. Connections between basic blocks can be represented graphically, which we refer to as a control flow graph (CFG). CFGs represent conditional flow of a program. 

\subsection{Verilog Netlist}
A Verilog-netlist encodes a hardware-design (e.g., an $n$-bit Adder, a registered \texttt{and}-gate) in terms of a device's available programmable logic resources (e.g., DSPs, LUTs, etc.). We will refer to a device's available programmable logic resources as device-specific primitives. In our work, we consider PLD netlist extraction processes which recover designs and represent them as a Verilog-netlist. We provide an example in Fig.~\ref{fig:dsn_verilog}. In Fig.~\ref{fig:dsn_verilog}(a) a registered-AND-gate is behaviourally described with Verilog. The same design is presented in Fig.~\ref{fig:dsn_verilog}(b); however, it has been encoded in terms of device-specific primitives.

\begin{figure*}[h]
    \centering
    \begin{subfigure}[t]{0.45\textwidth}
        \centering
        \begin{minted}[
        fontsize=\footnotesize,
        ]{verilog}
module reg_and (clk, a, b, out);
    input a,b,clk;
    output reg out;
    always @(posedge clk) begin
        out <= a & b;
    end
endmodule
        \end{minted}        
        \caption{}
    \end{subfigure}
    \begin{subfigure}[t]{0.45\textwidth}
        \centering
        \begin{minted}[
        fontsize=\footnotesize,
        ]{verilog}
module reg_and (clk, a, b, out);
  input a, b, clk;
  output out;
  wire a_inv;
  FDR fdr (.C(clk), .D(b), .R(a_inv), .Q(out));
  INV inv (.I(a), .O(a_inv));
endmodule
        \end{minted}        
        \caption{}
    \end{subfigure}
    \caption{An example of (a) behaviourally-described hardware design in Verilog and (b) the same design described using device-specific primitives.}
    \label{fig:dsn_verilog}
\end{figure*}
\section{vlang Approach}
We present \texttt{vlang}, a LLVM-frontend to compile a Verilog netlist into LLVM's IR. \texttt{vlang}'s compilation process for a Verilog netlist is outlined in Fig.~\ref{fig:vlang}. In this process, a Verilog netlist is first parsed into a software model. The software-model holds information regarding the primitives (e.g., LUTs, AND-Gates, Flip-Flops, etc.) in the supplied design, as well as the connections between primitives. During compilation, \texttt{vlang} replaces each primitive in the software-model with an LLVM-function. Replacement LLVM-functions are behavioural descriptions of the device-specific primitives. Once all primitives have been \textit{replaced} with LLVM-functions, \texttt{vlang} determines the order-of-execution for all of the LLVM-functions through a scheduling algorithm; the explicit parallel structures in hardware need to be mapped into sequentially executed code\footnote{\texttt{vlang} does not attempt to create multi-threaded programs; this is left as future work.}. After scheduling has completed, \texttt{vlang} will \textit{connect} LLVM-functions by assembling a set of temporary registers in the LLVM-IR to act as the wiring between modules. Lastly, \texttt{vlang} will produce a C-header file which outlines the calling-convention for this compiled design. The following subsections provide further detail on \texttt{vlang}'s compilation process.

\subsection{Mapping Primitive Modules to LLVM}

Typical PLD-fabric consists of programmable elements such as look-up tables (LUTs), flip-flops (FFs), digital-signal processing blocks (DSPs) and on-chip memories (RAMs). Routing resources are required to \textit{route} information to-and-from these hardware-blocks. A Verilog netlist describes the hardware design (which has been configured on the PLD) by structurally representing programmable-hardware resources (i.e., primitives) as Verilog \texttt{module}s. Connections between primitives are explicitly made through the use of \texttt{wire}s \cite{thomas2008verilog}. For each primitive in the design, \texttt{vlang} will construct an LLVM-function, which we describe below.

\subsection{Inputs and Outputs}
\begin{figure*}[!b]
    \centering
    \begin{subfigure}[t]{0.45\textwidth}
        \centering
        \begin{minted}[
        fontsize=\footnotesize,
        ]{verilog}
module example(a, b, c);
    input   [31:0] a;
    input   [31:0] b;
    output  [31:0] c;
    //== structural description.
endmodule
        \end{minted}        
        \caption{}
    \end{subfigure}
    \begin{subfigure}[t]{0.45\textwidth}
        \centering
        \begin{minted}[
        fontsize=\footnotesize,
        ]{llvm}
define void @example(i32 %a, i32 %b, i32* %c) {
    ; Translated Verilog module here..
    ;
    ;
    ret void
}
        \end{minted}        
        \caption{}
    \end{subfigure}
    \caption{An example of how \texttt{vlang} translates \texttt{input/output} nets to LLVM. (a) is an example of a Verilog module to be translated in LLVM-IR. (b) is the result.}
    \label{fig:input_output_example}
\end{figure*}

\texttt{vlang} remaps a primitive's \texttt{input}s into LLVM-software datatypes (e.g., \texttt{i1}, \texttt{i32}, \texttt{i43}) which match the bit-widths of the inputs\footnote{Both LLVM and Verilog support variable-sized bitwidth datatypes.} Since a Verilog module may have more than one output bus/net, \texttt{vlang} remaps output nets to pointers (sized to match the datatype-width of the net) as this provides the ability to obtain more than one output from a software function. Fig.~\ref{fig:input_output_example},  demonstrates this remapping, where Fig.~\ref{fig:input_output_example}(a) provides the original Verilog module and Fig.~\ref{fig:input_output_example}(b) provides the LLVM-function declaration generated by \texttt{vlang}. 

\subsection{Primitives}

Primitives on PLD fabric can either be combinational or clock triggered. We map both classes of primitives to LLVM-functions.

\subsubsection{Combinational Primitives}

Combinational primitives such as \{\texttt{AND}, \texttt{XOR}, \texttt{OR}, ...\} gates or LUTs, can be mapped directly to LLVM instructions (e.g., an \texttt{AND} gate is replaced with the \texttt{and} instruction). However, replacing the LUT combinational primitive is not as trivial. To map a LUT into an LLVM-function, the following information must be provided: (a) the size of the LUT (i.e., the number of inputs to the LUT) and (b) the LUT-mask. Using the LUT-mask, the sum-of-products formula is constructed, therefore, a sequence of \texttt{and} and \texttt{or} LLVM-instructions can be used to construct the behaviour of the LUT.

\subsubsection{Flip-Flops}
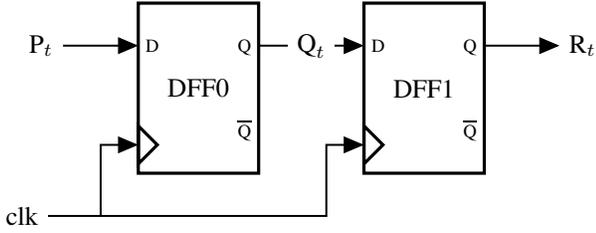
\begin{figure}[tbh!]
    \centering

    \begin{tikzpicture}[>=triangle 45]
    \def\N{1}  
    
    \foreach \m in {0,...,\N}
    \node [shape=dff] (DFF\m) at ($ 3*(\m,0) $) {DFF\m};
    
    \def\p{0}  
    \foreach \m in {1,...,\N} { 
    \draw [->] (DFF\p.Q) -- (DFF\m.D) node[midway,fill=white] {$\text{Q}_{t}$};
    \global\let\p\m
    }
    
    \draw [<-] (DFF0.D) -- +(-1,0) node [anchor=east] {$\text{P}_t$} ;
    \draw [->] (DFF\N.Q) -- +(1,0) node [anchor=west] {$\text{R}_t$};
    
    \path (DFF0) +(-2cm,-1.7cm) coordinate (temp)
    node [anchor=east] {clk};
    \foreach \m in {0,...,\N}
    \draw [->] (temp) -| ($ (DFF\m.CLK) + (-5mm,0) $) --(DFF\m.CLK);
    \end{tikzpicture}

    \caption{Motivating example illustrating the need for two global-variables for the LLVM-function equivalent.}
    \label{fig:dff_example}
\end{figure}
Flip-flops are clock-triggered primitives which provide storage capabilities. Replacement LLVM-functions must be able emulate a flip-flop. Consider the circuit in Fig.~\ref{fig:dff_example}, where two D-Flip Flops are \textit{chained} together. At the next clock edge, data at the D inputs of the flip-flops will be \textit{stored} and displayed at the output: DFF1 will display $\text{Q}_{t}$ at it's output, and DFF0 will output $\text{P}_t$. However, if we were to sequentially update the flips flops from left-to-right (i.e., DFF0 is updated, then DFF1), DFF1 would improperly update to output $\text{P}_t$. Unfortunately, a naive replacement LLVM-function would lead to the above error, since \texttt{vlang} produces sequentially executed LLVM-IR. Therefore, the replacement LLVM-function must preserve the parallel-functionality of the flip-flop. We detail how to update LLVM-described flip-flops. The function signature of the replacement LLVM-function for a flip-flop is depicted in Fig.~\ref{fig:dff_funcsig}.

\begin{figure}[tbh!]
    \centering
        \begin{minted}[
        fontsize=\footnotesize,
        ]{llvm}
define void @dff(i1 %d, i1 %clk, i1* %q) {
    ; Translated Verilog module here..
    ret void
}
        \end{minted}  
    \caption{The LLVM-function signature of a D-flip-flop.}
    \label{fig:dff_funcsig}
\end{figure}

Each flip-flop replacement must have \textit{two} global variables, GV1 and GV2. GV1 holds the state of an incoming changes. GV2 holds the current state of the flip-flop, which will then be returned at the end of the function and then updated with the contents of GV1. Additionally, the flip-flop should only change state on a clock-edge.  However, LLVM lacks the notion of \textit{edge-triggered} logic. To capture a clock-edge, the function must be able to \textit{recall} the previous state of the clock, and then use this memory as a reference against the current clock-signal. \texttt{XOR}-ing the recalled and current clock-signal will dictate if there was a change in the clock. Then, inspecting the current clock level will dictate if this was a positive edge, or negative edge shown in Fig.~\ref{fig:llvmclocklogic}. Other flip-flop types (e.g., asynchronous reset, clock-enabled) can also be constructed as an LLVM function, by modifying this base-design. 

\begin{figure}[tbh!]
	\centering
	\begin{minted}[
	fontsize=\footnotesize,
	]{llvm}
define i32 @dff(i1 %d, i1 %clk, i1* %q) {	
    %load.clk = load i32* @out_1.clk_reg
    %clkstate = xor i32 %clk, %load.clk
    %clkstate1 = and i32 %clk, %clkstate
    %clk_hi = icmp eq i32 %clkstate1, 1
    ;... if hi, save d.
    ; Remember the state of the clock
    store i1 %clk, i1* @dff.clk_reg
    ;...
}
	\end{minted}  
	\caption{Clock edge detection logic in LLVM.}
	\label{fig:llvmclocklogic}
\end{figure}

\subsubsection{Other Primitives}

We currently do not support DSP and RAM blocks due the large number of configuration possibilities. Instead, if DSPs or RAM blocks are used in the original hardware, we map these to LUT and Flip-Flop elements. Adding support for these primitives is left as future work. We summarize the mapping between Verilog primitives and LLVM Datatypes in Table.~\ref{tab:mapping}.

\begin{table*}[tbh!]
	\centering
	\caption{Mapping between Verilog primitives and LLVM structures.}
	\label{tab:mapping}
	\resizebox{\textwidth}{!}{%
		\begin{tabular}{l|c|c|c|c|c}
			\toprule
			\textbf{Verilog}       	& Primitive Declaration    	& Inputs and Wires 	& Outputs   & Logic Gates        & Other Primitives\\\hline
			\textbf{LLVM}    		& Function Declaration    	& Variables 		& Pointers  & Logic Instructions & LLVM Functions \\
			\bottomrule
		\end{tabular}
	}
\end{table*}

\subsection{Module Execution Schedule}

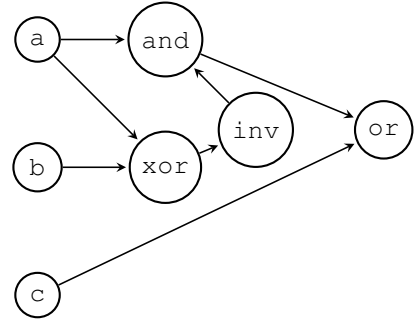
\begin{figure}[tbh]
    \centering
    \begin{tikzpicture}[
            > = stealth, 
            shorten > = 1pt, 
            auto,
            node distance = 1.7cm, 
            semithick 
        ]

        \tikzstyle{every state}=[
            draw = black,
            thick,
            fill = white,
            minimum size = 2mm
        ]

        \node[state] (i0)  {\texttt{a}};
        \node[state] (i1) [below of=i0] {\texttt{b}};
        \node[state] (i2) [below of=i1] {\texttt{c}};
        \node[state] (v1) [right of=i0] {\texttt{and}};
        \node[state] (v2) [below right of=v1] {\texttt{inv}};
        \node[state] (v3) [right of=i1] {\texttt{xor}};
        \node[state] (v4) [right of=v2] {\texttt{or}};

        \path[->] (i0) edge node {}(v1);
        \path[->] (i1) edge node {}(v3);
        \path[->] (i2) edge node {}(v4);
        \path[->] (i0) edge node {}(v3);
        \path[->] (v2) edge node {}(v1);
        \path[->] (v3) edge node {}(v2);
        \path[->] (v1) edge node {}(v4);

    \end{tikzpicture}
    \caption{An example of a device-specific netlist encoded as a data flow graph.}
    \label{fig:dfg}
\end{figure}

All primitives mapped into LLVM functions may only be executed sequentially and must preserve the functionality of the original hardware design. However, Verilog-netlists are a form of dataflow-graphs (DFGs) which expose explicit parallelism between operators. An example of this encoding is pictured in Fig.~\ref{fig:dfg}, where the logic equation:
\begin{equation}\label{eq:logicexample}
(\texttt{a}\land(\sim(\texttt{a}\oplus\texttt{b}))) \lor \texttt{c}    
\end{equation}
is encoded as a DFG. Additionally, Verilog-netlists may contain feedback (e.g., a linear-shift feedback register). \texttt{vlang} must determine an ordering of the nodes in the dataflow graph which preserves the data flow between nodes (or synonymously, primitives). Typically, a topological ordering would suffice if the DFG was guaranteed to be acyclic \cite{kahn1962topological}. Since Verilog-netlist DFGs may contain feedback paths, a topological ordering cannot be computed unless feedback paths in the graph can be handled.

Feedback loops in Verilog netlists can be divided into two classes: (a) sequential feedback and (b) combinational feedback. If a feedback edge passes through a clock-triggered element which saves state (i.e., a flip-flop), then this will be referred to as sequential feedback. Otherwise, combinational feedback occurs if all primitives along the feedback edge are a form of combinational logic. Our tool does not support combinational logic loops.

To schedule the execution of primitives, the corresponding DFG must be acyclic. We present a methodology to (a) identify feedback-edges (i.e., cycles) from flip-flops in a DFG, (b) temporarily modify the DFG to be acyclic by removing feedback, (c) schedule the acyclic DFG, and then reintroducing feedback edges to restore the functionality of the original circuit. 

\subsection{Breaking Sequential Feedback Loops}

First, \texttt{vlang} must identify if there are cycles in the provided DFG. Recall, the DFG is a directed graph. \texttt{vlang} employs Tarjan's Strongly-Connected Components algorithm \cite{tarjan1972depth} to the DFG, to discover any strongly-connected components (SCC), thereby discovering cycles. The vertices in each discovered SCC are inspected to \textit{ensure} there exists at-least one flip-flop (as this permits valid feedback paths). If there are no flip-flops in any of the strongly-connected components, \texttt{vlang} will abort compilation. Otherwise, the flip-flops within the SCC must be modified to \textit{break} the cycle/SCC (providing an acyclic graph). 
\begin{algorithm}[tbh]
\caption{Cycle Removal Algorithm}\label{alg:cycle_removal}
\begin{algorithmic}[1]

\Function{CycleRemoval}{$G$}
\State $C \gets \texttt{TarjanSCC(}G\texttt{)}$ \Comment{List of SCCs}
\State $R \gets \texttt{newKeyValueTable()}$
\If{\texttt{HasCombinationalSCC(}C\texttt{)}}
\State \texttt{abort()}
\EndIf
\For{$c \in C$}
    \For{$v \in c$}
        \If {$\texttt{IsaFlipFlop(}v\texttt{)}$}
            \For{$U \in v\texttt{.getModulesCon2In()}$}
                \If{$U \in c$}
                    \State $R\texttt{[}v\texttt{].append(}U\texttt{)}$
                    \State $v\texttt{.removeInput(}U\texttt{)}$
                \EndIf
            \EndFor
        \EndIf
    \EndFor 
\EndFor
\State \Return $R$
\EndFunction
\end{algorithmic}
\end{algorithm}
Our procedure for flip-flop modification is outlined in Algorithm~\ref{alg:cycle_removal}. For each flip-flop within an SCC, the modules which provide input (i.e., drivers) to the flip flop are inspected to see if the module is within the SCC. If a flip-flop's driver is within the vertex set of the SCC, the \textit{edge} (i.e., wire) connecting them is cut. The flip-flop driver pair is noted down in $R$, a key-value data structure, as these modules will need to be reconnected after scheduling has occurred.

\subsection{ASAP Scheduling}

Once Algorithm~\ref{alg:cycle_removal} has completed, \texttt{vlang} generates an ILP formulation for an LP solver to serialize the data-flow graph. We used the system of difference constraints ILP formulation from~\cite{cong2006efficient}, which will provide a solution for an as-soon-as-possible (ASAP) execution schedule. We used the LP solver, \texttt{lpsolve}~\cite{berkelaar2004lpsolve}, to compute the schedule. As an example, if we serialized the DFG-encoding of Equation~\ref{eq:logicexample} using the ILP, the order of execution would be: $(\texttt{xor}, \texttt{inv}, \texttt{and}, \texttt{or})$. Optimizations to the serialization process could be made by adjusting the SDC-constraint formulation; we have left this as future work. 

However, with Algorithm~\ref{alg:cycle_removal}, flip-flops will be scheduled to be connected to what \textit{it} drives. For example, if a flip-flop drives an \texttt{and} gate, the flip-flop will be scheduled to execute \textit{just-before} the \texttt{and} gate. However, the execution schedule of the modified flip-flops are not guaranteed to have the correct data driving it's inputs. 
\begin{algorithm}[tbh]
\caption{\texttt{vlang}'s ASAP Scheduling}\label{alg:scheduling}
\begin{algorithmic}[1]
\Function{Schedule}{}
\State $G_{\text{circuit}}(E,V)$\Comment{This is the DFG of the circuit} 
\State $R \gets \textsc{CycleRemoval(}G\textsc{)}$ \Comment{From Alg.~\ref{alg:cycle_removal}}
\State $S \gets \texttt{ASAP(}G\texttt{)}$ \Comment{Computed Serialized Schedule}
\For{$v \in S$}
    \If{$v \in R$}
        \State $v_\text{copy} \gets v\texttt{.MakeCopy()}$
        \State $v_\text{copy}\texttt{.disableClock()}$
        \State $S.\texttt{replace(}v,v_\text{copy}\texttt{)}$
    \EndIf
\EndFor
\For{\{$v, (U_0,U_1, ...)\} \in R$}
    \State $D \gets v\texttt{.getInputs()}$
    \State $T \gets \texttt{GetLatestStart(}D_0, D_1, ... ,U_0, U_1, ...\texttt{)}$
    \State $S\texttt{.insertAfter(}v, T\texttt{)}$
\EndFor
\State \Return $S$
\EndFunction
\end{algorithmic}
\end{algorithm}
Therefore, modifications to the schedule must be made to allow for the flip-flops to (a) propagate data to the necessary consumers and (b) permit the updating of the flip-flops by reattaching their original drivers. We outline this procedure in Algorithm~\ref{alg:scheduling}. For each modified flip-flop, we make a copy of the flip flop and set the clock to zero (thereby disabling the flip-flop). This allows the consumers of the affected flip-flops to obtain the data at the correct time. However, the affected flip-flops also need to be updated at the correct time. Recall, \texttt{vlang} had stored the cut drivers from the flip-flops in a key-value data structure, $R$. To restore updates to the flip-flops, the original copy of the flip-flop is inserted into the schedule. The insertion point is just-after the \textit{latest} start-time of all of it's original inputs. This restores updates to the flip-flop and guarantees the driver's data to be arriving at the flip-flop at the correct time. 

Once the schedule has been generated and modified, the original DFG has been serialized. \texttt{vlang} will iterate through the schedule, and begin inserting \texttt{call} instructions to the the corresponding primitive-modules. Connections between modules are made with local variables, which are known as \texttt{alloca} instructions in LLVM. \texttt{vlang} uses the connection-information collected during it's parsing stage to determine how may \textit{temporary} registers (in the LLVM-IR sense) should be instantiated to simulate data-transfer using wires. If a subsection of a bus is set or read, \texttt{vlang} will use \texttt{and/or} logic to set or read from the sub-selected bus.

\subsection{Cleaning-Up}

Upon completion of schedule-implementation, \texttt{vlang} produces a C header-file which provides the function-signature to \textit{call} the top-level modules as a software function.

\section{Experimental Setup and Results}

A variety of experiments are performed with \texttt{vlang} to showcase the ability to preserve portability of Verilog-netlist. First, we demonstrate \texttt{vlang}'s ability to produce a software-executable which preserves the functionality and cycle accuracy of the Verilog netlist. The software executable is linked to a C-testbench program. The results of the C-testbench are compared against three state-of-the-art hardware simulators: Verilator~\cite{snyder2017verilator}, Icarus Verilog~\cite{williams2006icarus} and Modelsim ~\cite{graphics2007modelsim}. We then explore the usage of \texttt{vlang} as a front-end for HLS tools, by exploring if the HLS-generated circuit from \texttt{vlang}'s LLVM-IR is functionally correct. We also explore the achieved performance and area from the high-level synthesis of \texttt{vlang}'s LLVM-IR, and compare against the original design's performance and area. To conduct our studies, we explore a number of examples from OpenCore~\cite{opencore}.

\subsection{OpenCore Benchmarks}

OpenCore is an online repository of hardware-designs represented in Verilog or VHDL. A number of designs from OpenCores have been selected to evaluate \texttt{vlang}'s ability to handle: (a) combinational-logic circuits, and (b) sequential-logic circuit with the possibility of feedback. 
\subsubsection{Combinational Logic Circuits}
\begin{itemize}
    \item[] \textbf{adder}: is a ripple-carry adder which has (a) three inputs, two 16-bit numbers and a 1-bit carry-in and (b) two outputs, one 16-bit sum, and a 1-bit carry-out.
    \item[] \textbf{bcdadder}: implements the addition of two 4-bit numbers and a 1-bit carry-in, and produces a 4-bit sum as a binary-coded-decimal with a 1-bit carry-out.
    \item[] \textbf{divide}: performs the division of two 32-bit inputs, and outputs a 32-bit quotient.
    \item[] \textbf{mod3}: calculates a 3-bit remainder of an 8-bit input in modulo-3 space. 
    \item[] \textbf{popcount32}: determines the number of one's in a 32-bit input and is output onto a 5-bit bus.
\end{itemize}

\subsubsection{Benchmarks with Clock-Triggered Elements}
\begin{itemize}
    \item[] \textbf{addertree}: implements an adder tree for five 16-bit inputs, where each pair of inputs are added and registered. The last registered stage is output to a 16-bit bus. 
    \item[] \textbf{andreg}: computes the logical-and between two single-bit inputs, registers and then returns the result.
    \item[] \textbf{gcd}: calculates the greastest-common-divisor (GCD) between two 32-bit numbers, storing the result in a 32-bit register. This benchmark has feedback.
\end{itemize}

We selected these benchmarks to provide a variety of tests with respect to: (a) complexity of the design, (b) clock-triggered behaviour with and without feedback, and (c) a variety of applications. However, these benchmarks are described \textit{behaviourally}; \texttt{vlang} can only compile a Verilog-netlist into LLVM-IR. To simulate this environment, we assume that is it possible to retrieve designs from Xilinx's Kintex-7 FPGAs in the form of Verilog netlists. Using \texttt{xst} from Xilinx's ISE design suite \cite{xilinx2015design}, and targeting a Kintex-7 device (\texttt{xc7k70t-1fbg484}), we use \texttt{xst} to synthesize a behavioural design from our OpenCores benchmark suite into the Kintex-7's FPGA-fabric, thereby producing a device-specific netlist. The device-specific netlist can be represented as Verilog-netlist using Xilinx's \texttt{netgen}.

\subsection{Results: Functional Correctness and Simulation Comparison}

 Recall that \texttt{vlang} produces an LLVM-IR program which reproduces the functionality of a Verilog-netlist, and thereby, the original design. Simultaneously, \texttt{vlang} also produces a C-header file to permit other programs to reference and use this program. The \texttt{vlang} produced LLVM-IR can be linked into a C program using the LLVM framework \cite{lattner2008llvm}. Once linked, the corresponding LLVM-IR can either be executed just-in-time \cite{mogensen2009basics} or compiled to a binary. To evaluate \texttt{vlang}, we compile the test-bench and \texttt{vlang}-produced LLVM-IR to a binary using \texttt{llc} and \texttt{gcc}~\cite{griffith2002gcc}. Each C test-bench tests a large subset of all possible test-cases, providing an avenue to validate correctness. The number of unique test-cases issued for each OpenCore benchmark is detailed in Table~\ref{tab:testcases}. We also compare \texttt{vlang}'s simulation results and times against two open-source tools, Icarus Verilog \cite{williams2006icarus} and Verilator \cite{snyder2017verilator} and a commercial simulation tool, \texttt{Modelsim}. All simulators were provided with the same test-cases. For all of the OpenCore benchmarks, Icarus Verilog, Modelsim and \texttt{vlang} produced the correct results. However, Verilator is unable to produce correct results for the \texttt{addertree} and \texttt{gcd} benchmarks. This is due to a software-bug in Verilator, where certain feedback paths in the Verilog netlist are not handled correctly.

 \begin{table*}[tbh!]
     \centering
 \caption{Number of unique test-cases performed on the OpenCore benchmarks.}
 \label{tab:testcases}
\begin{tabular}{cccccc|ccc}
\toprule
& \textbf{adder} & \textbf{bcdadder} & \textbf{divide} & \textbf{mod3} & \textbf{popcount32} & \textbf{addertree} & \textbf{andreg} & \textbf{gcd} \\
            \cmidrule(lr){2-9}   
\# Testcases & 2359296 & 256 & 65536 & 256 & 1048576 & 625 & 8 & 262144 \\
Icarus Verilog Passed? & Y & Y  & Y & Y  & Y & Y  & Y & Y \\
Verilator Passed? & Y & Y  & Y & Y  & Y & \textbf{N}  & Y & \textbf{N} \\
Modelsim Passed? & Y & Y  & Y & Y  & Y & Y & Y & Y \\
\texttt{vlang} Passed? & Y & Y  & Y & Y  & Y & Y  & Y & Y \\
\bottomrule
\end{tabular}
 \end{table*}
 

\pgfplotstableread[row sep=\\,col sep=&]{
    benchmark   & iv                    & v         & vlang     & msim\\
    adder       & 67.2                  & 6.87      & 7         & 261\\
    divide      & 54.43                 & 28.43     & 1.53      & 19 \\
    popcount32  & 38.97                 & 5.72      & 2.67      & 43\\
    addertree   & 452.87                & 55.83     & 34.65     & 909\\
    gcd         & 5124.00               & 56.84     & 27.75     & 99 \\
    }\simulationresults

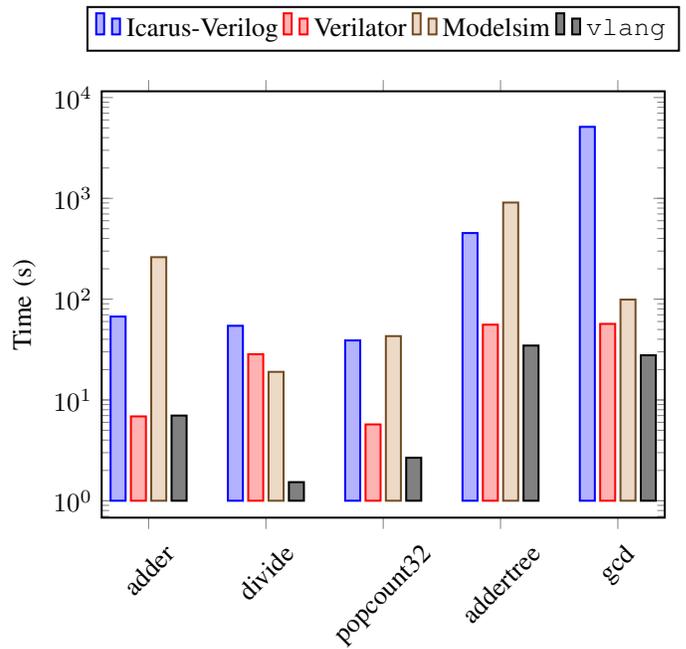
\begin{figure}
    \centering
    \begin{tikzpicture}
        \begin{axis}[
                ybar,
                ymode=log,
                bar width=.20cm,
                width=0.5\textwidth,
                height=.4\textwidth,
                legend style={at={(0.5,1.2)},
                anchor=north,legend columns=-1},
                symbolic x coords={adder, divide, popcount32, addertree, gcd},
                xtick=data,
                ylabel={Time (s)},
                ylabel style={yshift=-0.2cm},
                xticklabel style={rotate=45},        
            ]
            \addplot table[x=benchmark,y=iv]{\simulationresults};
            \addplot table[x=benchmark,y=v]{\simulationresults};
            \addplot table[x=benchmark,y=msim]{\simulationresults};
            \addplot table[x=benchmark,y=vlang]{\simulationresults};
            \legend{Icarus-Verilog, Verilator, Modelsim, \texttt{vlang}}
        \end{axis}
    \end{tikzpicture}
    \caption{Comparison of Icarus Verilog \cite{williams2006icarus}, Verilator \cite{snyder2017verilator} and \texttt{vlang}'s simulation time (in Log Scale) for the OpenCore benchmarks. Results for \textbf{bcdadder}, \textbf{mod3} and \textbf{andlatch} are not shown due to fast simulation time (on the order of $10^{-3}$ seconds) }
    \label{fig:simres}
\end{figure}

 Additionally, we compare the execution time for each testbench with all three simulators. The results are outlined in Fig.~\ref{fig:simres}. This figure does not include results for \textbf{bcdadder}, \textbf{mod3} and \textbf{andreg}, as simulation times were on the order of $10^{-3}$ seconds. 
For a majority of the benchmarks, \texttt{vlang} is able to achieve a faster simulation time, compared to Icarus Verilog, Verilator and Modelsim. This is expected, as \texttt{vlang} is a compiled simulator \cite{zivojnovic1996compiled}; these styles of simulators are known to provide faster execution time compared to event-driven simulators (Icarus Verilog, Verilator and Modelsim are event driven simulators) \cite{wang1991lecsim}. Overall \texttt{vlang} is $\sim37.3 \times$ faster than Icarus Verilog, $\sim3.5\times$ faster than Verilator and $\sim12.3 \times$ faster than Modelsim. Although the focus of this paper is not on simulation, we were able to present \texttt{vlang}'s ability to: (a) produce an executable which reproduces the functionality of a Verilog-netlist with an LLVM-IR program and (b) be used a simulation tool, which is competitive in terms of simulation time and is cycle accurate.

\subsection{Results: Porting between PLDs}

We evaluate \texttt{vlang}'s portability by supplying \texttt{vlang}-produced LLVM-IR to LegUp, an open-source high-level synthesis tool from the University of Toronto \cite{canis2011legup}. Using LegUp, we compile the \texttt{vlang}-produced LLVM-IR to a Verilog design file, and target two devices, a Xilinx Kintex-7 device (\texttt{xc7k70t-1fbg484}) and an Intel Stratix V device (\texttt{5SGXEA7N2F45C2}), to study if the functionality of these designs are portable. With the Xilinx device, we also measure performance and area differences: the original design files (i.e., not the Verilog-netlists) are compiled with Xilinx's ISE v14.7 and targeted for the Kintex-7 device. For Xilinx devices, area is gauged in terms of slices and number of LUTs and performance is measure in terms of $F_\text{max}$. For the combinational circuits in our OpenCore Benchmark suite, we used the worst-case estimated logic delay as our measure of $F_\text{max}$. The post place-and-route (PnR) area-and-performance metrics for the original design files are outlined in Table~\ref{tab:kintex7perfnarea}.
\begin{table}[tbh!]
    \centering
    \caption{ISE's Post PnR Kintex-7 Performance and Area metrics for OpenCore benchmarks.}
    \label{tab:kintex7perfnarea}
    \resizebox{0.5\textwidth}{!}{%
        \begin{tabular}{lcccc}
            \toprule
            & \multicolumn{3}{c}{ \textbf{Area} } & \textbf{Performance} \\
            \cmidrule(lr){2-4}   
            \cmidrule(lr){5-5}
            \textbf{Benchmark}   & \textbf{Slices}  & \textbf{LUTs} & \textbf{Flip-Flops} & \textbf{$F_\text{max}$} (MHz) \\
            \cmidrule(lr){1-5}
            \textbf{adder}       & 4    & 16 & - & 136.44 \\
            \textbf{bcdadder}    & 4    & 12 & - & 143.32 \\
            \textbf{divide}      & 477  & 1347 & - & 11.86 \\
            \textbf{mod3}        & 2    & 3 & - &  164.10 \\
            \textbf{popcount32}  & 19    & 53 & - & 103.61 \\
            \cmidrule(lr){1-5}
            \textbf{addertree}   & 16   & 64 & 64 & 671.14 \\
            \textbf{andreg}    & 2    & 4 & 1 & 709.723 \\
            \textbf{gcd}         & 41  & 135 & 93 & 429.73 \\
            \bottomrule
        \end{tabular}
    }
\end{table}

To compare the HLS-generated circuits (from \texttt{vlang}'s LLVM-IR) to the original design files, we compiled the design's Verilog-netlists with \texttt{vlang}. The \texttt{vlang}-produced IR was then supplied to LegUp. LegUp was set produce Xilinx-amenable Verilog, with a 1 ns clock-period. The HLS-generated circuit then underwent ISE's compilation flow, again targeting the Kintex-7 device, with a 1 ns clock-period. The results are tabulated in Table~\ref{tab:kintex7perfnarealegup}. Using our C-test-benches along with \texttt{vlang}-produced IR, we confirmed the functionality of the Verilog-netlist from the Kintex-7 was successfully ported back to the same device using LegUp's test-bench generation flow~\cite{canis2011legup}. However, supplying \texttt{vlang}'s LLVM-IR to an HLS tool may affect the overall area of HLS-generated design as shown in Table~\ref{tab:kintex7perfnarealegup}. This is demonstrated for all benchmarks, where the number of Slices and LUTs increase. As a general trend, the increase in LUTs follows the complexity of the original design: for example, the \textbf{gcd} benchmark uses 135 Slice-LUTs in the original design and the \texttt{vlang}-mapped design uses 5952 LUTs, whereas the \textbf{adder} benchmark uses 16 LUTs in the original design and \texttt{vlang}-mapped design uses 370. Additionally, the \textbf{divide} benchmark is not route-able on the Kintex-7 device (the design uses 97\% of the available slices on this particular device).  The increase in area is primarily attributed to the HLS-tool's ability to produce hardware designs from \texttt{vlang}'s LLVM-Functions. The LLVM-function replicates the behaviour of a low-level device-primitive; there is no guarantee that the HLS tool will remap these LLVM-functions into these same primitive. Instead, the HLS tool may proceed to use additional logic resources to implement the LLVM-function. Additionally, extra logic from: (a) the HLS-compiler will be introduced to the original design (e.g., introducing an FSM-controller to enable switching between functions), thereby adding additional area costs and (b) from duplicateds flip-flops on feedback paths. Another notable difference: the combinational circuits have transformed into sequential circuits. This is expected since HLS-tools will infer a clock-signal (which is propagated throughout the entire design). This allows HLS tools to insert register stages into the design, to improve of upon the performance (i.e., breaking-up the critical path, pipe-lining, etc.). This can be desirable, especially if the combinational circuit need not be combinational with the new device. In our case all combinational benchmarks had a significant improvement in $F_{\text{max}}$. If cycle-latency is a sensitive parameter for design portability, the current flow will not be suitable: the additional register stages may affect the circuit's cycle-latency. LegUp (or any LLVM-based HLS tool) could be modified to reflect the original design's clock usage (or lack thereof) in \texttt{vlang}'s IR, but this is left as future work. Introducing register stages into combinational logic also prevents Quartus from inferring the entire logic function into a LUT: several LUTS and register would need to be used to infer the behaviour of this function.

\begin{table}[tbh!]
    \centering
    \caption{ISE's Post PnR Kintex-7 Performance and Area metrics for OpenCore benchmarks (files regenerated with LegUp HLS from \texttt{vlang}). \text{\sffamily X} indicates the design was unroutable.}
    \label{tab:kintex7perfnarealegup}
    \resizebox{0.5\textwidth}{!}{%
        \begin{tabular}{lcccc}
            \toprule
            & \multicolumn{3}{c}{ \textbf{Area} } & \textbf{Performance} \\
            \cmidrule(lr){2-4}   
            \cmidrule(lr){5-5}
            \textbf{Benchmark}      & \textbf{Slices}  & \textbf{LUTs} & \textbf{Flip-Flops} & \textbf{$F_\text{max}$} (MHz) \\
            \cmidrule(lr){1-5}
            \textbf{adder}          & 214       & 370   & 389 & 709.72 \\
            \textbf{bcdadder}       & 178       & 318   & 280 & 709.72 \\
            \textbf{divide}         & 9841      & 37,219 & 36,627 & \text{\sffamily X} \\
            \textbf{mod3}           & 90        & 200   & 151 & 709.72 \\
            \textbf{popcount32}     & 319       & 636   & 583 & 709.72 \\
            \cmidrule(lr){1-5}
            \textbf{addertree}      & 1334      & 2289  & 2175 & 594.88 \\
            \textbf{andreg}         & 11        & 29    & 18 & 709.72 \\
            \textbf{gcd}            & 3625      & 5952  & 8908 & 472.14 \\
            \bottomrule
        \end{tabular}
    }
\end{table}

Lastly, we confirm the portability of \texttt{vlang}'s LLVM-IR by using LegUp HLS and targeting a Stratix V device. LegUp was set to target a 1 ns clock-period. The LegUp produced hardware design was then compiled with Quartus Prime 18.0 Standard Edition, where we set a target clock-period of 1 ns, and targeted the Stratix V device \texttt{5SGXEA7N2F45C2}. We use the Stratix V ALM (adaptive logic module) count as our area metric. We characterize performance through $F_\text{max}$. Again, using our C-test-benches we used LegUp's test-bench generation flow to verify the functionality of the Verilog-netlist from the Kintex-7 was successfully ported to a Stratix V device. Table~\ref{tab:stratixvperfnarea} tabulates the post place-and-route performance and area metrics for the Intel Device. 

\begin{table}[tbh!]
    \centering
    \caption{Quartus's Post PnR Stratix V Performance and Area metrics for the high-level synthesis of \texttt{vlang}-compiled OpenCore benchmarks.}
    \label{tab:stratixvperfnarea}
    \resizebox{0.5\textwidth}{!}{%
        \begin{tabular}{lcccc}
            \toprule
            & \multicolumn{3}{c}{ \textbf{Area} } & \textbf{Performance} \\
            \cmidrule(lr){2-4}   
            \cmidrule(lr){5-5}
            \textbf{Benchmark}   & \textbf{ALMs}  & \textbf{Registers} & \textbf{ALUT} & \textbf{$F_\text{max}$} (MHz)\\
            \cmidrule(lr){1-5}
            \textbf{adder}          & 57    & 141 & 74 & 508.65 \\
            \textbf{bcdadder}       & 52    & 130 & 71 & 543.48   \\
            \textbf{divide}         & 18153    & 37519 & 36349 & 307.88 \\
            \textbf{mod3}           & 23    & 53 & 29 & 714.29 \\
            \textbf{popcount32}     & 54    & 140 & 69 & 399.04  \\
            \cmidrule(lr){1-5}
            \textbf{addertree}      & 407   & 428 & 625 & 381.24 \\
            \textbf{andreg}         & 32    & 58 & 54 & 554.32    \\
            \textbf{gcd}            & 3079  & 6135 & 5258 & 317.06 \\
            \bottomrule
        \end{tabular}
    }
\end{table}

We note on the performance and area characteristics on the HLS-generated circuits for both Xilinx and Intel devices. Although there is no direct area comparison between Intel and Xilinx devices, we compare the number of LUTs used in Xilinx device (from Table~\ref{tab:kintex7perfnarea}) to the number of ALUTs used in the Intel device. As before, we observe an increase in area: this is primarily attributed to the HLS-tool's inability to recognize the device-specific primitives as LLVM-functions. Lastly, the achieved $F_{\text{max}}$ for the combinational benchmarks when using \texttt{vlang} with LegUp HLS improve upon the baseline. Again, this is due to the HLS tool's ability to (a) insert a clock signal to the design, and (b) insert register stages to break-up the critical path. Referring to Table~\ref{tab:kintex7perfnarealegup} and Table~\ref{tab:stratixvperfnarea}, there is a performance difference between the designs on the Intel device versus the Xilinx device. From our investigation, we discovered that Quartus had purposed many LUTs as route-throughs introducing additional logic delay, thereby affecting the  $F_{\text{max}}$. This differs from Xilinx's ISE, where the majority of designs utilized routing tracks versus LUTs as route-throughs, thereby providing a higher $F_{\text{max}}$ overall. From our experiments, \texttt{vlang}'s IR was able to port designs originally intended for a Kintex-7 Xilinx FPGA to an alternative device technology, an Intel Stratix V device. We also demonstrated \texttt{vlang}'s LLVM-IR can be used as a software executable and can be integrated with other software-ecosystems.

\section{Related Work} \label{sec:related}

We present a list of works related to: (a) Verilog-to-C compilation procedures, (b) a  (c) high-level synthesis technologies

\subsection{Verilog-to-C/C++ Compilers}

\subsubsection{Verilog to C/C++ Compilers for Hardware Simulation}
Open-source tools \texttt{v2c} \cite{mukherjee2016v2c}, \texttt{icarus} \cite{williams2006icarus} and \texttt{verilator} \cite{snyder2017verilator} compile hardware description languages (i.e., Verilog and VHDL) to C/C++ for verification purposes. However, these tools do not present HLS-synthesizable C/C++ code.

\subsubsection{Mapping Verilog to C/C++ for High-Level Synthesis}

Bomberi et al. explored raising the level of abstraction of hardware designs for: (1)  generating and exploring designs with an HLS compiler \cite{bombieri2013method}, and (2) using IP cores as standalone system software \cite{bombieri2015methodology}. This work was able to \textit{recover} the IP block specification for system-level design, and enable the derivation of more optimized implementations through HLS.

Mahapatra and Schafer identify a similar issue: older designs intended for a target architecture may suffer when porting HDL~\cite{mahapatra2019veriintel2c}. Additionally, their may be a need to optimize and port RTL cores~\cite{mahapatra2019veriintel2c,bombieri2013method}. The authors provide a tool, veriIntel2c, to compile a subset of Verilog to a synthesizable set of C  \cite{mahapatra2019veriintel2c}. The compiled code is then explored for pareto-optimal designs by applying different code modifications to the C program. The authors built on this work in \cite{mahapatra2019optimizing}, where they optimize their Verilog to C process.

In these works, only either (a) a small subset of Verilog can be retargeted to C/C++ programs~\cite{mahapatra2019veriintel2c} or (b) an event scheduler-kernel is required to \textit{simulate} a circuit in software~\cite{bombieri2013method,bombieri2015methodology}. Our work is able to produce a software executable that is scheduled at compile time, while handling all gate-level representations in HDL.

\subsection{LLVM-Based High-Level Synthesis Tools}

High-level synthesis is a compilation flow which compiles a high-level language (e.g., C/C++) to a hardware description language (e.g., Verilog, VHDL)~\cite{coussy2008high}. A number of algorithmic processes are required to map a high-level language to a hardware description language. The program will undergo a number of static and dynamic program analyses to either restructure the program to expose parallelism and make the program amenable for hardware. The restructured program will then undergo allocation, scheduling and binding. Generally, HLS tools take advantage of pre-exising compilation frameworks, such as LLVM or GCC~\cite{canis2011legup,pilato2013bambu}. Our work produced LLVM-IR, to be suitable for usage with HLS compilers built within LLVM frameworks.

\section{Conclusion}

This paper proposes a compilation method which transforms Verilog netlists into LLVM-IR to allow portability of these designs across a variety of computer architectures (e.g., $\mu$Processors, FPGAs, etc) by harnessing the LLVM-framework~\cite{lattner2004llvm}. Our method was implemented as an LLVM-frontend, which compiles the Verilog netlist to LLVM's IR. Our frontend, \texttt{vlang}, is able to produce an LLVM-IR program which preserves the functionality of the Verilog-netlist and is cycle accurate. Due to this ability, \texttt{vlang} can be used a verification tool. Lastly, \texttt{vlang}'s LLVM-IR is HLS-friendly. A number of experiments were performed with \texttt{vlang}. First, \texttt{vlang}'s LLVM-IR was validated by using brute-force comparisons against the original Verilog design. This simultaneously demonstrated \texttt{vlang}'s ability to function as a Gate-level simulator and it's ability to retain the exact functionality and cycle accuracy of a hardware design as a software executable.  Lastly, we evaluated \texttt{vlang}'s LLVM-IR for it's usage in HLS processes. We demonstrated designs originally on a Xilinx device were successfully migrated to (a) the original Xilinx device and (b) an Intel FPGA. We commented on the HLS-generated hardware design's performance and area metrics against (a) the original design's performance and area metrics on the original device and (b) against the original design compiled to the Intel device.


\bibliographystyle{./IEEEtran}
\bibliography{./conf}

\end{document}